\documentclass{article}
\usepackage[preprint]{spconf}
\usepackage{amsmath,graphicx}

\usepackage{multirow}
\usepackage{amssymb}
\usepackage[bookmarks=false]{hyperref}
\usepackage{textcomp}

\copyrightnotice{\copyright\ 2022 IEEE}
\toappear{To appear in {\it Proc.\ ICASSP 2022, May 07-13, 2022, Virtual}}

\title{unsupervised audio-caption aligning learns correspondences between individual sound events and textual phrases \vspace{-6pt}}
\name{Huang Xie, Okko R\"as\"anen, Konstantinos Drossos, Tuomas Virtanen \vspace{-6pt}}
\address{Unit of Computing Sciences, Tampere University, Finland \vspace{-18pt}}
\begin{document}
    \setlength{\abovedisplayskip}{3pt}
    \setlength{\belowdisplayskip}{3pt}
    \maketitle
    \begin{abstract}
        We investigate unsupervised learning of correspondences between sound events and textual phrases through aligning audio clips with textual captions describing the content of a whole audio clip.
        We align originally unaligned and unannotated audio clips and their captions by scoring the similarities between audio frames and words, as encoded by modality-specific encoders and using a ranking-loss criterion to optimize the model.
        After training, we obtain clip-caption similarity by averaging frame-word similarities and estimate event-phrase correspondences by calculating frame-phrase similarities.
        We evaluate the method with two cross-modal tasks: audio-caption retrieval, and phrase-based sound event detection (SED).
        Experimental results show that the proposed method can globally associate audio clips with captions as well as locally learn correspondences between individual sound events and textual phrases in an unsupervised manner.
    \end{abstract}
    \vspace{-10pt}

    \section{Introduction}
    \label{sec:introduction}
    \vspace{-6pt}

    Cross-modal learning, which aims at processing and relating information across multimodal data (e.g., audio, image and text), has received increasing attention recently.
    In this paper, we focus on cross-modal learning between audio and natural language descriptions, i.e., audio-text cross-modal learning.
    Generally, natural language allows the description of acoustic information and the versatile modeling of sound relationships in ways that are understandable by humans.
    Automated interpretation of audio data with natural language has great potential in real-world applications, such as audio retrieval, acoustic monitoring, and human-computer interaction.

    Recent audio research that deals with cross-modal learning across audio and text modalities includes audio-text retrieval~\cite{Elizalde2019Cross}, automated audio captioning~\cite{Drossos2017Automated, Xu2019ACRNN}, and audio question answering~\cite{Fayek2020Temporal}.
    Audio-text retrieval concerns matching of audio and text pairs.
    Elizalde et al.~\cite{Elizalde2019Cross} associated audio with text by jointly learning representations of audio and text with a siamese network.
    In automated audio captioning~\cite{Drossos2017Automated, Xu2019ACRNN}, captions were automatically generated for audio clips to summarize sound events contained in them.
    Audio question answering investigates acoustic reasoning through answering textual questions pertaining to audio clips.
    The pioneering work~\cite{Fayek2020Temporal} predicted answers across predefined values by learning joint representations of audio and questions.

    The aforementioned studies~\cite{Elizalde2019Cross, Drossos2017Automated, Xu2019ACRNN, Fayek2020Temporal} mainly focus on information fusion across audio and text modalities, for example, learning joint audio-text representations.
    However, the relationships between audio and text elements are rarely investigated.
    For example, audio clips consists of sound events, which are described by textual phrases in human-annotated captions~\cite{Kim2019AudioCaps, Drossos2020Clotho}.
    It is not fully understood what information the above cross-modal learning methods learn when matching whole audio clips and their captions.

    Inferring the latent relationships between audio and text elements is a key to automated interpretation of audio data using textual descriptions.
    The problem of finding relationships between elements from different modalities is referred to as cross-modal alignment, which has been widely studied in the language and vision communities~\cite{Baltrusaitis2019Multimodal}.
    For example, given an image and a written or spoken caption, cross-modal alignment aims to associate the image regions with the caption\textquotesingle s words or phrases~\cite{Harwath2016Unsupervised, Hu2019Multilevel, Wehrmann2020Adaptive, Xu2020Cross, Harwath2020Jointly, Khorrami2021Evaluation, Wang2021Align}.
    These studies have shown the potential for learning element-wise correspondences across different modalities via cross-modal alignment.
    Particularly, Harwath et al.~\cite{Harwath2020Jointly} demonstrate that visual objects and spoken words can be jointly discovered from raw images and speech signals by aligning image pixels with speech waveform in an unsupervised manner.
    In contrast to supervised alignment, which requires labeled aligned training data, unsupervised alignment can explore cross-modal correspondences without resorting to expensive data annotations.

    We propose an unsupervised audio-text aligning method to investigate the correspondences between sound events and textual phrases within clip-caption pairs.
    Previous work~\cite{Xu2021Text} explored event-phrase correspondences with a supervised approach, where timestamps and textual phrases of sound events were provided for clip-caption pairs during model training.
    In contrast, we utilize audio clips and captions that are temporally unaligned and unannotated, aside from knowing which of them belong together.
    We obtain clip-caption similarities by aggregating frame-word similarities.
    The proposed method is evaluated on two cross-modal tasks: audio-caption retrieval and phrase-based SED\@.
    Experimental results show that the proposed method can learn global associations between audio clips and captions as well as local correspondences between individual sound events and textual phrases.

    \vspace{-10pt}

    \begin{figure}[!t]
        \centering
        \includegraphics[width=1.0\linewidth]{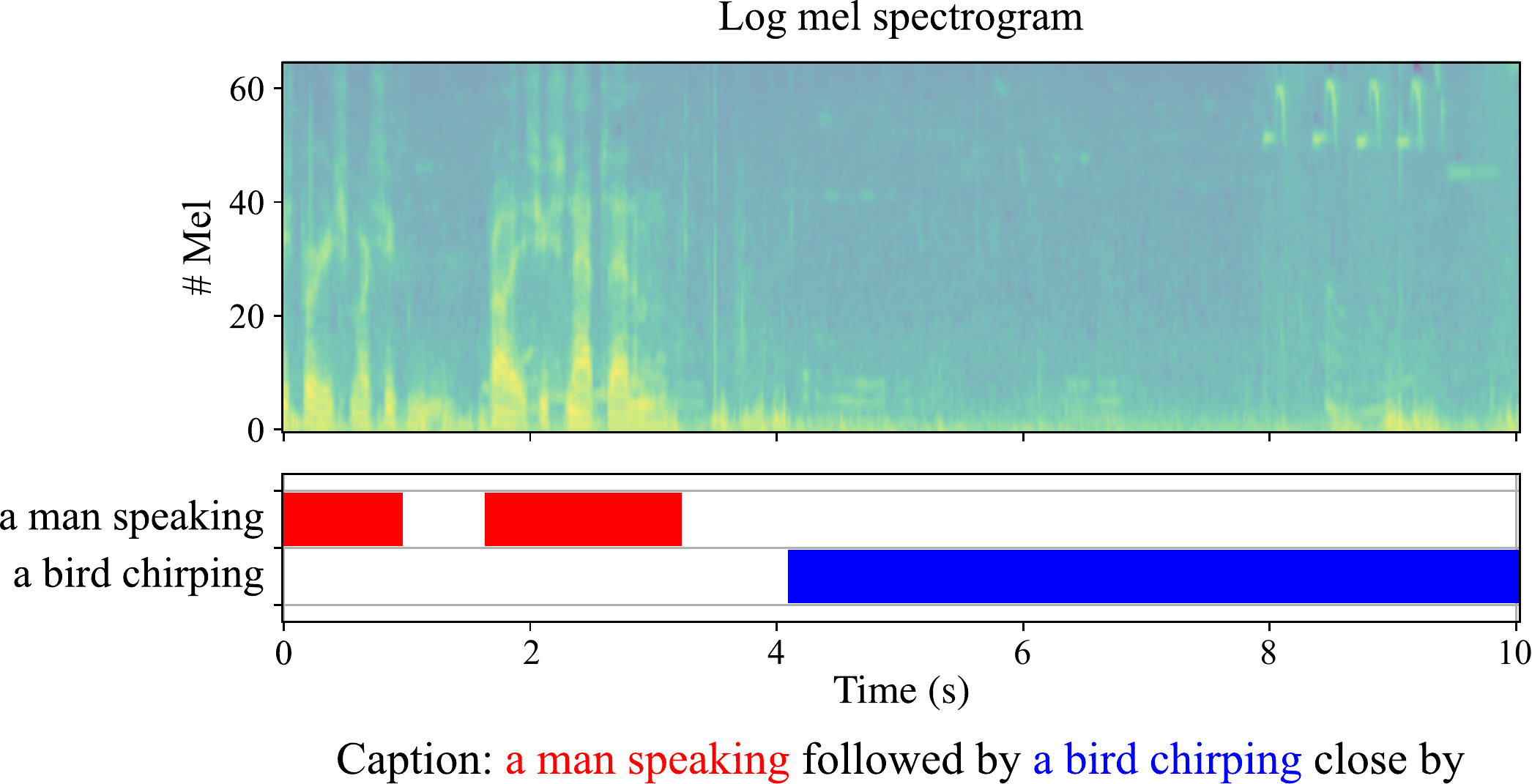}
        \vspace{-18pt}
        \caption{An example of the temporal presences of two sound events in an audio clip and their corresponding descriptive phrases in the clip\textquotesingle s caption. The audio clip is illustrated with its log mel spectrogram, and the temporal activity of each sound event is shown with colored bars.}
        \label{fig:sample_illustration}
        \vspace{-10pt}
    \end{figure}
    \begin{figure*}[!t]
        \centering
        \includegraphics[width=1.0\textwidth]{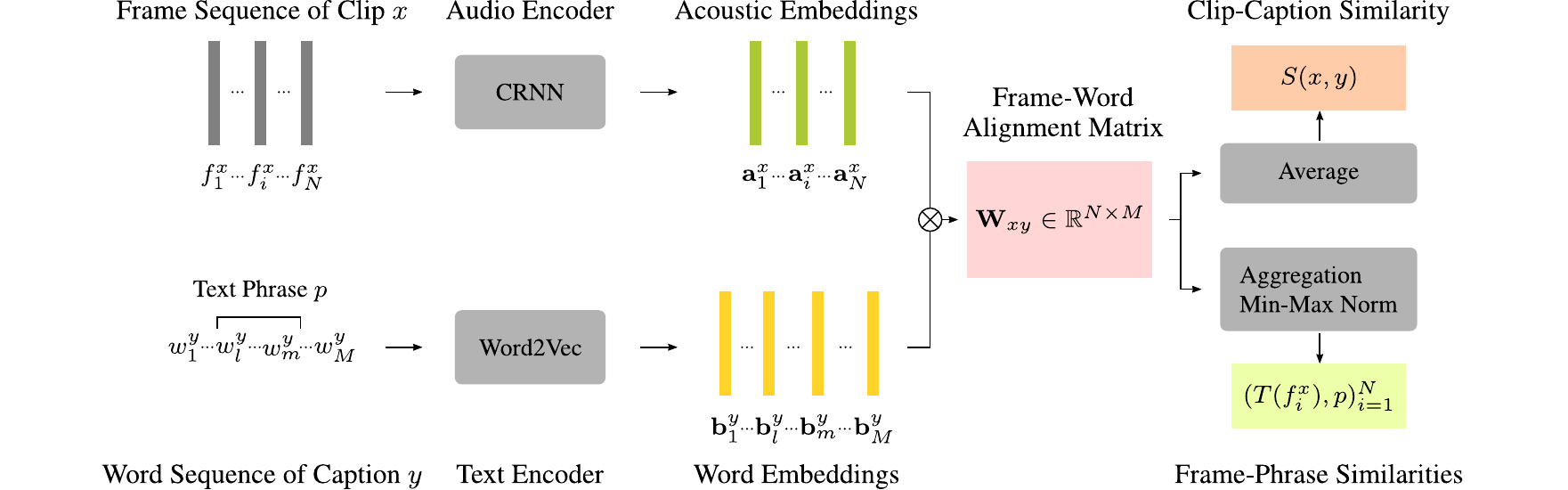}
        \vspace{-18pt}
        \caption{The proposed audio-text aligning framework.}
        \label{fig:aligning_framework}
        \vspace{-10pt}
    \end{figure*}

    \section{Proposed Method}
    \label{sec:proposed-method}
    \vspace{-6pt}

    In this section, we present the proposed unsupervised audio-text aligning method for learning event-phrase correspondences and clip-caption similarities.
    Figure~\ref{fig:sample_illustration} shows an example clip-caption pair containing two sound events and their corresponding textual phrases which are targeted in this study.
    \vspace{-24pt}

    \subsection{Audio-Text Aligning Framework}
    \label{subsec:audio-text-aligning-framework}
    \vspace{-3pt}

    Inspired by previous works~\cite{Harwath2020Jointly, Xu2021Text}, we propose an audio-text aligning framework with two input encoders: one for audio, and the other for text, as illustrated in Figure~\ref{fig:aligning_framework}.
    Alignment between audio clip $x$ and its caption $y$ involves processing and relating audio information in $x$ and text information in $y$, as well as mapping together similarities of their acoustics and lexical semantics.
    We split $x$ into a sequence of $N$ audio frames $(f_{i}^{x})_{i=1}^{N}$ and represent $y$ by a sequence of $M$ words $(w_{j}^{y})_{j=1}^{M}$.
    A textual phrase $p$ is defined by a sub-sequence $(w_{l}^{y}, \dots, w_{m}^{y})$ of $y$, which ranges from the $l$-th to $m$-th words.

    The audio encoder extracts frame-wise acoustic embeddings $(\mathbf{a}_{i}^{x})_{i=1}^{N} \in \mathbb{R}^{300}$ from $x$, and the text encoder converts $y$ into a sequence of word-specific semantic embeddings $(\mathbf{b}_{j}^{y})_{j=1}^{M} \in \mathbb{R}^{300}$.
    A frame-word alignment matrix $\mathbf{W}_{xy} \in \mathbb{R}^{N \times M}$ of $x$ and $y$ is obtained by scoring the similarity between each audio frame and word pair.

    Following~\cite{Harwath2020Jointly}, we define the similarity of the $i$-th audio frame $f_{i}^{x}$ and the $j$-th word $w_{j}^{y}$ by the dot product of their embeddings $\mathbf{a}_{i}^{x}$ and $\mathbf{b}_{j}^{y}$
    \begin{equation}
        \label{eq:embedding_similarity}
        sim(f_{i}^{x}, w_{j}^{y}) = dot(\mathbf{a}_{i}^{x}, \mathbf{b}_{j}^{y}).
    \end{equation}
    To associate audio clips with captions, we rely upon the positive frame-word similarities and discard the negatives.
    Therefore, the alignment matrix $\mathbf{W}_{xy}$ can be written as
    \begin{equation}
        \label{eq:alignment_matrix}
        \mathbf{W}_{xy} = (s_{ij} (x,y)) \in \mathbb{R}^{N \times M},
    \end{equation}
    where $s_{ij} (x,y) = \max (0, sim(f_{i}^{x}, w_{j}^{y}))$ is the trimmed similarity of the $i$-th audio frame $f_{i}^{x}$ and the $j$-th word $w_{j}^{y}$.
    With $\mathbf{W}_{xy}$, we calculate the global clip-caption similarity $S(x, y)$ between $x$ and $y$ by averaging the matrix elements
    \begin{equation}
        \label{eq:global_similarity}
        S(x, y) = \dfrac{1}{N M} \sum_{i=1}^{N} \sum_{j=1}^{M} s_{ij} (x,y).
    \end{equation}

    We train the audio-text aligning framework by optimizing a ranking-based criterion~\cite{Bromley1993Signature}, such that audio clips and captions that belong together are more similar than mismatched clip-caption pairs.
    Specifically, across a batch of $K$ clip-caption pairs $\{(x_{k},y_{k})\}_{k=1}^{K}$, where $y_{k}$ is a caption pertaining to an audio clip $x_{k}$, we randomly select imposter clip $\hat{x}_{k}$ and imposter caption $\hat{y}_{k}$ for each clip-caption pair $(x_{k},y_{k})$.
    Then, we calculate the widely used sampling-based triplet loss~\cite{Harwath2016Unsupervised, Harwath2020Jointly, Harwath2020Learning, Chrupaa2021Visually, Khorrami2021Can}
    \begin{equation}
        \label{eq:triplet_loss}
        \begin{split}
            loss = \dfrac{1}{K} \sum_{k=1}^{K} & [ \max (0, S(x_{k},\hat{y}_{k}) - S(x_{k},y_{k}) + \eta)      \\
            & + \max (0, S(\hat{x}_{k},y_{k}) - S(x_{k},y_{k}) + \eta) ],
        \end{split}
    \end{equation}
    where $\eta$ is a margin hyper-parameter.
    We fix $\eta$ to one.
    \vspace{-12pt}

    \subsection{Phrase-based SED}
    \label{subsec:phrase-based-sed}
    \vspace{-3pt}

    The phrase-based SED refers to detecting sound events within an audio clip based on textual phrases describing them.
    It can be employed as a proxy task for investigating what the model presented in the previous section learns.
    Sound events are detected by examining the similarities between audio frames and textual phrases.
    We obtain frame-phrase similarities by aggregating $\mathbf{W}_{xy}$ along words contained in $p$.
    The frame-phrase similarity $T(f_{i}^{x}, p)$ is calculated by aggregating the trimmed frame-word similarities $\{s_{il} (x,y), \dots, s_{im} (x,y)\}$
    \begin{equation}
        \label{eq:local_similarity}
        T(f_{i}^{x}, p) = aggregation (s_{il} (x,y), \dots, s_{im} (x,y)).
    \end{equation}
    Then, min-max normalization is applied to the frame-phrase similarities $(T(f_{i}^{x}, p))_{i=1}^{N}$ across all the audio frames of one audio clip.
    For event detection, event $e$ is predicted to be present in $f_{i}^{x}$ if $T(f_{i}^{x}, p)$ is above a detection threshold $\gamma$.
    \vspace{-12pt}

    \subsection{Audio Encoder}
    \label{subsec:audio-encoder}
    \vspace{-3pt}

    We use a convolutional recurrent neural network (CRNN)~\cite{Xu2019ACRNN} as the audio encoder.
    It consists of five convolution blocks, followed by a bidirectional gated recurrent unit (BiGRU).
    Each convolution block includes an initial batch normalization, a convolutional layer with padded $3 \times 3$ convolutions, and a LeakyReLU activation with a slope of $-0.1$.
    After the first, third, and fifth convolution blocks, one L4-Norm subsampling layer is used to reduce the temporal dimension of each block\textquotesingle s output by a factor of four.
    A dropout layer with a rate of $0.3$ is placed between the last L4-Norm layer and the BiGRU.
    Lastly, an up-sampling operation is applied to ensure the final output has the same temporal dimension as the CRNN input.

    The CRNN audio encoder takes 64-dimensional log mel-band energies as input.
    Each audio clip is split into 40 ms Hanning-windowed frames with a hop length of 20 ms.
    Then, 64 log mel-band coefficients are extracted from each frame.
    A sequence of 300-dimensional acoustic embeddings are generated for each audio clip.
    The final acoustic embeddings are normalized with L2 normalization.
    \vspace{-12pt}

    \subsection{Text Encoder}
    \label{subsec:text-encoder}
    \vspace{-3pt}

    We utilize Word2Vec (Skip-gram model)~\cite{Mikolov2013Efficient} as the text encoder.
    Word2Vec is a two-layer fully-connected neural network, which learns word embeddings that are good at predicting surrounding words in a sentence or a document.
    For the sake of simplicity, we adopt a publicly available pre-trained Word2Vec~\cite{Word2Vec_online}, which is trained on Google News dataset.
    It consists of 300-dimensional word embeddings for roughly three million case-sensitive English words and phrases.
    The Word2Vec text encoder converts textual descriptions into sequences of semantic word embeddings word by word.
    The final word embeddings are normalized with L2 normalization, and then fixed in our experiments.
    \vspace{-10pt}

    \section{Experiments}
    \label{sec:experiments}
    \vspace{-6pt}

    In this section, we evaluate the proposed method with the AudioGrounding dataset~\cite{Xu2021Text} on two cross-modal tasks: audio-caption retrieval, and phrase-based SED\@.
    \vspace{-12pt}

    \subsection{Dataset}
    \label{subsec:dataset}
    \vspace{-3pt}

    The original AudioGrounding dataset~\cite{Xu2021Text} consists of 4,590 10-second audio clips, 4,994 descriptive captions, and 13,985 sound event phrases.
    It is split into three sets: a training set with 4,489 clips, a validation set with 31 clips, and a test set with 70 clips.
    All audio clips are drawn from YouTube videos, and their corresponding captions are sourced from Audiocaps~\cite{Kim2019AudioCaps}.
    One human-annotated descriptive caption is provided for each clip in the training set while five captions are provided for each in the validation and test sets.
    Sound event phrases are extracted automatically from captions, and human annotators are invited to merge the extracted phrases that correspond to the same sound event and provide the timestamps of each sound event~\cite{Xu2021Text}.

    In this work, we collect audio clips of the AudioGrounding dataset from their original YouTube videos.
    Because of unavailable YouTube videos, we have 4,253 clips for the training set, 30 clips for the validation set, and 67 clips for the test set.
    The statistics of our downloaded version for the AudioGrounding dataset are shown in Table~\ref{tab:Statistics_AudioGrounding}.
    \vspace{-12pt}

    \begin{table}[!t]
        \centering
        \begin{tabular}{c||c|c|c}
            \hline
            \bfseries Split & \#\bfseries Clips & \#\bfseries Captions & \#\bfseries Event phrases \\
            \hline
            Train           & 4253              & 4253                 & 11732                     \\
            \hline
            Val             & 30                & 150                  & 439                       \\
            \hline
            Test            & 67                & 335                  & 1118                      \\
            \hline
        \end{tabular}
        \caption{Statistics of the downloaded dataset.}
        \label{tab:Statistics_AudioGrounding}
        \vspace{-6pt}
    \end{table}

    \subsection{Experimental Setup}
    \label{subsec:experimental-setup}
    \vspace{-3pt}

    We train the aligning framework with batches of 32 clip-caption pairs in the training set for at most 100 epochs, and monitor the loss~\eqref{eq:triplet_loss} on the validation set during the training process.
    An Adam optimizer with an initial learning rate of $0.001$ is adopted to optimize the training process.
    The learning rate is reduced by a factor of ten once the validation loss does not improve for five epochs.
    Training is terminated by early stopping with ten epochs.
    \vspace{-12pt}

    \subsection{Evaluation}
    \label{subsec:evaluation}
    \vspace{-3pt}

    We evaluate the proposed method with the test set in the AudioGrounding dataset on the tasks of audio-caption retrieval and phrase-based SED\@.

    \textbf{Audio-Caption Retrieval}.
    This task aims to retrieve instances that are most similar to a given instance from one modality to the other modality, such as retrieving captions pertaining to an audio clip.
    The task serves to provide a single high-level metric, which captures how well the aligning framework has learned to bridge audio and text modalities at the whole-clip and whole-caption levels.
    We experiment on audio-to-caption and caption-to-audio retrieval with a clip-wise negative sampling approach, where the correct target caption/clip for a given query clip/caption is always among 29 negative randomly sampled captions/clips.
    The number of negative samples is determined by the validation size, which only has 30 unique clips.
    For each query probe in the test set, audio clips and captions are sorted by their clip-caption similarities~\eqref{eq:global_similarity}.
    Retrieval performance is measured in terms of recall at $k$ (R@\textit{k} with \textit{k} $\in \{1, 5\}$) averaged across all the audio clips, i.e., checking if the target clip/caption is always within top \textit{k} search results according to the clip-caption similarity sore.
    Theoretically, the chance levels are $1/{30}$ (around $0.03$) for R@1 and $1/{6}$ (around $0.17$) for R@5, respectively.
    To prevent randomness, we repeat the evaluation twenty times on the test set.

    \textbf{Phrase-based SED}.
    This task aims to detect sound events within audio clips using sound event phrases in the clips\textquotesingle ~captions.
    It serves to explore the ability of the aligning framework on learning correspondences between individual sound events and their textual phrases.
    Given an audio clip and a sound event phrase from its caption, we experiment with mean and max aggregations to calculate the frame-phrase similarities~\eqref{eq:local_similarity}, followed by a min-max normalization.
    Detection performance is measured with global frame-based metrics (precision, recall, and F1), which are calculated across all clips and sound event phrases.
    Random guessing is utilized as the baseline that generates frame-phrase similarities with random numbers from a uniform distribution on the interval $\left[ 0,1 \right)$.
    \vspace{-12pt}

    \subsection{Results and Analysis}
    \label{subsec:results-and-analysis}
    \vspace{-3pt}

    \begin{table}[!t]
        \centering
        \begin{tabular}{c|c||c|c}
            \hline
            \multicolumn{2}{c||}{\bfseries Model} & \bfseries Chance Levels & \bfseries Proposed \\
            \hline
            \hline
            \multirow{2}{*}{Audio2Caption} & R@1 & 0.03 & 0.21 $\pm$ 0.04 \\
            \cline{2-4}
            & R@5 & 0.17 & 0.65 $\pm$ 0.05 \\
            \hline
            \hline
            \multirow{2}{*}{Caption2Audio} & R@1 & 0.03 & 0.23 $\pm$ 0.04 \\
            \cline{2-4}
            & R@5 & 0.17 & 0.71 $\pm$ 0.04 \\
            \hline
        \end{tabular}
        \caption{Recall scores of audio-caption retrieval on the test set of the AudioGrounding dataset.}
        \label{tab:audio-caption-retrieval}
    \end{table}

    \textbf{Audio-Caption Retrieval}.
    The averages and standard deviations of recall scores with the proposed method are reported in Table~\ref{tab:audio-caption-retrieval}.
    In contrast to the theoretical chance levels, the proposed method achieves better recall scores, with R@1 over $0.20$ and R@5 over $0.65$.
    The experimental results show that the proposed method can associate audio clips with their corresponding captions, and vice versa.

    \begin{figure}[!t]
        \centering
        \includegraphics[width=1.0\linewidth]{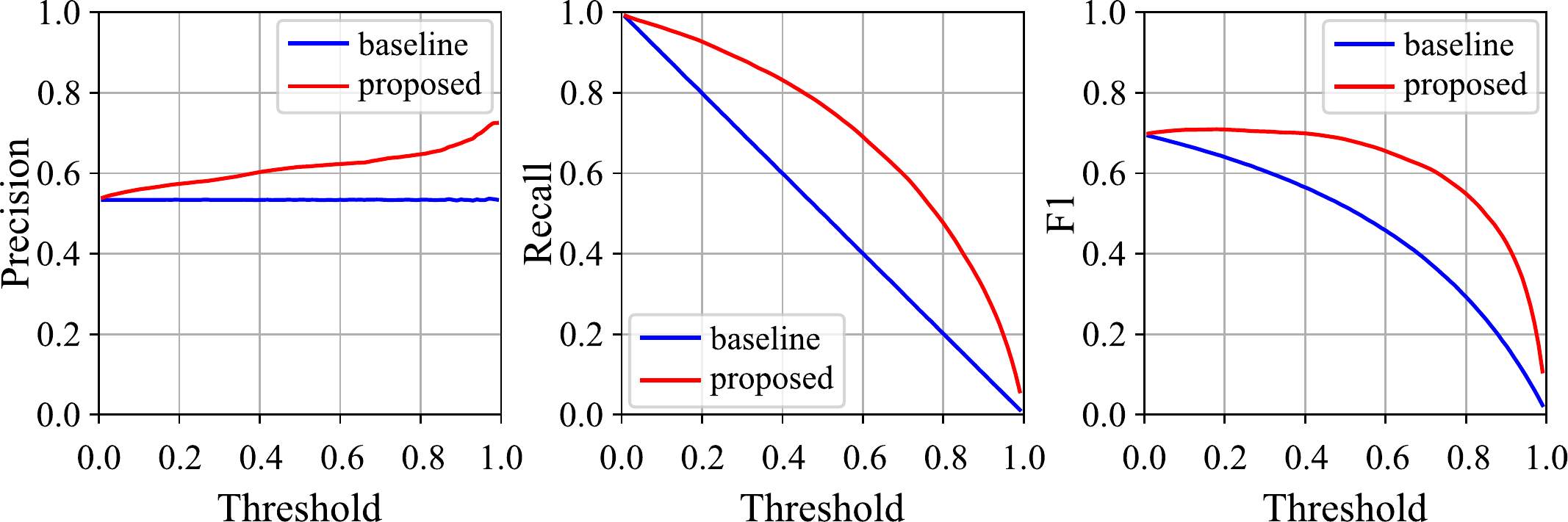}
        \vspace{-18pt}
        \caption{Frame-based metrics (precision, recall, and F1) against detection threshold for phrase-based SED on the test set of the AudioGrounding dataset.}
        \label{fig:sed_metrics}
        \vspace{-10pt}
    \end{figure}

    \textbf{Phrase-based SED}.
    The frame-based metrics (precision, recall, and F1) against detection threshold are illustrated in Figure~\ref{fig:sed_metrics}.
    The proposed method obtains similar results with either mean or max aggregations in~\eqref{eq:local_similarity}, and only the results from mean aggregations are reported.
    Overall, the proposed method achieves better detection performance than random guessing on the three metrics, regardless of the detection thresholds.
    Particularly, with a detection threshold of $0.5$, the proposed method obtains a precision of $0.62$, a recall of $0.77$, and a F1 of $0.68$ while random guessing has values of $0.53$, $0.50$, and $0.52$, respectively.
    The experimental results show that the proposed method can detect sound events from audio clips with their textual phrases.
    Additionally, we experiment with the proposed method by removing min-max normalization from calculating frame-phrase similarities, which results in a drastic reduction in the detection performance.

    \begin{figure}[!t]
        \centering
        \includegraphics[width=1.0\linewidth]{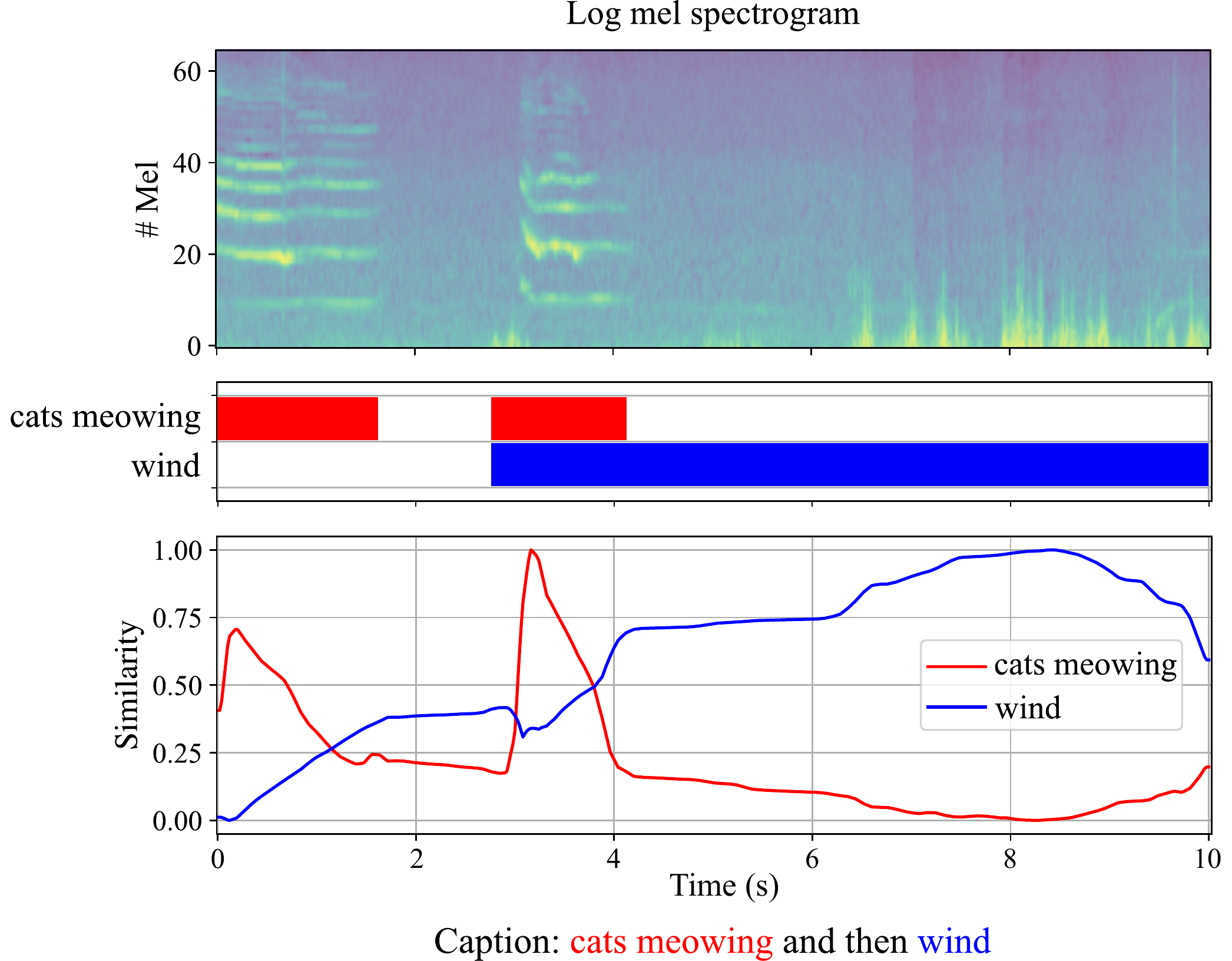}
        \vspace{-18pt}
        \caption{An example of learned frame-phrase similarities for two sound events and their textual phrases.}
        \label{fig:learned_similarities}
        \vspace{-10pt}
    \end{figure}

    An example of learned frame-phrase similarities for two sound events and their corresponding textual phrases is illustrated along time in Figure~\ref{fig:learned_similarities}.
    It shows that high similarities are learned only between sound events and their corresponding textual phrases.
    We conclude that the proposed method can learn event-phrase correspondences via unsupervised clip-caption aligning.
    \vspace{-10pt}

    \section{Conclusion}
    \label{sec:conclusion}
    \vspace{-6pt}

    We propose an unsupervised method to match sound events with textual phrases by aligning whole captions and audio clips.
    Audio clips and captions are aligned by scoring the frame-word similarities with their acoustic and semantic embeddings.
    We evaluate our proposed method with the AudioGrounding dataset~\cite{Xu2021Text} on two cross-modal tasks: audio-caption retrieval, and phrase-based SED\@.
    Experimental results show that the proposed method can learn global associations between audio clips and captions as well as local correspondences between individual sound events and textual phrases.
    As future work, we consider exploring large audio-text alignment datasets and fine-tuned semantic embeddings.
    \vspace{-20pt}

    \section{Acknowledgement}
    \label{sec:acknowledgement}
    \vspace{-6pt}

    The research leading to these results has received funding from Emil Aaltonen foundation funded project ``Kielen k\"aytt\"o structuroimattoman datan automaattiseen tulkintaan'' and Academy of Finland grant no. 314602.

    \bibliographystyle{IEEEbib}
    \bibliography{strings,refs}

\end{document}